\documentclass[final,5p]{elsarticle}
\biboptions{sort&compress}

\usepackage{amssymb}
\usepackage{bbm}
\usepackage{amssymb,graphicx}
\usepackage[intlimits]{amsmath}
\usepackage[small]{subfigure}
\usepackage{color}
\usepackage{array}
\usepackage{multirow}

\usepackage[colorlinks=true]{hyperref}
\hypersetup{
 linkcolor=blue,citecolor=blue,urlcolor=black}
 
\allowdisplaybreaks

\newcommand{\SU}{\mathit{SU}}
\newcommand{\U}{\mathit{U}}
\newcommand{\SO}{\mathit{SO}}

\newcommand{\osp}{\mathfrak{osp}}
\newcommand{\so}{\mathfrak{so}}
\newcommand{\su}{\mathfrak{su}}
\newcommand\eq[1]{eq.~(\ref{eq:#1})}
\newcommand{\ib}{{\bar{\imath}}}
\DeclareMathOperator{\tr}{tr} 
\newcommand{\dd}{\text{d}}
\newcommand{\iu}{\text{i}}
\newcommand{\gYM}{g_{\text{\tiny YM}}}
\newcommand{\CB}{\mathcal{B}}
\newcommand{\CF}{\mathcal{F}}
\newcommand{\CN}{\mathcal{N}}
\newcommand{\CO}{\mathcal{O}}
\newcommand{\CW}{\mathcal{W}}
\newcommand{\Ffbdy}{\mathfrak{f}^{\text{\tiny bdy}}}
\newcommand{\FFbdy}{\mathfrak{F}^{\text{\tiny bdy}}}
\newcommand{\Fhbdy}{\mathfrak{h}^{\text{\tiny bdy}}}

\journal{Physics Letters B}

\begin{document}

\begin{frontmatter}

\title{Superconformal Two-Point Functions of the Nahm Pole Defect\\
in ${\cal N}=4$ Super-Yang-Mills Theory}

\author{J.\ Baerman, A.\ Chalabi and C.\ Kristjansen}
\address{Niels Bohr International Academy, Niels Bohr Institute, Copenhagen University,\\ Blegdamsvej 17, 2100 Copenhagen \O{}, Denmark}
\ead{qwt632@alumni.ku.dk,adam.chalabi@nbi.ku.dk,kristjan@nbi.dk}

\begin{abstract}
We derive a manifestly superconformal expression for the leading-order two-point functions of all single trace chiral primary operators in 4d ${\cal N}=4$ super-Yang-Mills theory with a co-dimension one Nahm pole defect.
Notably, our result involves only a finite number of superblocks that correspond to 1/2-BPS representations in the defect channel. 
The derivation builds on a non-trivial exact result for traces of fuzzy spherical harmonics in combination with powers of $\su(2)$ generators, which leads to a remarkably compact expression for the bulk-to-defect couplings.
Apart from being interesting in its own right as a closed-form expression at finite $N$, the final result is an essential prerequisite for applying the defect superconformal bootstrap programme to this model.
\end{abstract}


\end{frontmatter}

\section{Introduction}
\label{sec:intro}

The Nahm pole boundary conditions of 4d ${\cal N}=4$ super-Yang-Mills (SYM) theory constitute a highly tract-able defect conformal field theory (DCFT) with maximal supersymmetry. 
Due to a non-trivial background field configuration, the computation of non-trivial correlation functions requires only few Wick contractions. 
Furthermore, the theory is amenable to several exact methods such as integrability~\cite{deLeeuw:2015hxa,Buhl-Mortensen:2015gfd,Buhl-Mortensen:2017ind,Gombor:2020kgu,Komatsu:2020sup,Kristjansen:2020mhn,Gombor:2020auk}, supersymmetric localisation~\cite{Wang:2020seq,Komatsu:2020sup} and the defect conformal bootstrap programme~\cite{McAvity:1995zd,Liendo:2012hy,Billo:2016cpy,deLeeuw:2017dkd}.
Whereas the application of the former two techniques can be said to have reached maturity with e.g.\ an exact expression for one-point functions of protected operators to all loop orders and for finite $N$~\cite{Komatsu:2020sup} as well as an exact expression for the all-loop planar one-point functions of non-protected operators~\cite{Gombor:2020kgu,Komatsu:2020sup,Gombor:2020auk}, the application of the latter method is far from having reached its full potential. 

After the modern analysis of refs.~\cite{Liendo:2012hy,Billo:2016cpy} the defect conformal bootstrap programme has gained significant momentum. 
Not least the Lorentzian inversion formula of~\cite{Caron-Huot:2017vep} and its defect versions~\cite{Lemos:2017vnx,Isachenkov:2018pef,Liendo:2019jpu} have provided novel tools for handling DCFTs by allowing to reconstruct defect data from bulk data alone and vice versa. 
Among models treated with these tools are the ordinary 3d and long-range Ising models~\cite{Liendo:2019jpu,Behan:2023ile} and more general $O(N)$-like models~\cite{Bertucci:2022ptt,Gimenez-Grau:2022ebb,Bianchi:2022sbz,Bianchi:2023gkk} as well as line defects in 4d ${\cal N}=2$ superconformal field theories (SCFTs)~\cite{Gimenez-Grau:2019hez} and ${\cal N}=4$ SYM theory~\cite{Barrat:2020vch,Barrat:2021yvp}.
 
Applications of the conformal bootstrap to co-dimen\-sion one defects or boundaries, however, are fewer. 
Although being simpler in certain aspects, the co-dimension one case involves additional subtleties that remain to be fully understood.
E.g.\ the boundary inversion formulas of ref.~\cite{Mazac:2018biw} require data from both bulk and boundary channels, unlike their higher co-dimension analogues.

In this paper we determine for the Nahm pole in $\CN=4$ SYM theory all defect channel data in the bulk chiral primary two-point function to leading order in the Yang-Mills (YM) coupling.
We derive a manifestly superconformal expression, which decomposes in terms of the boundary superblocks derived in ref.~\cite{Liendo:2016ymz}. 
Remarkably, it involves only a finite number of boundary superblocks that correspond to 1/2-BPS multiplets being exchanged in the defect channel.
Their coefficients have a compact closed form.

Now that the leading-order data of one channel is known, this model would be amenable to analysis via an appropriate inversion formula if one can be found.
Furthermore, it is a rich toy model with perturbative control for exploration of the analytic superconformal boundary bootstrap.
The first steps in applying the boundary conformal bootstrap to the Nahm pole defect were taken in ref.~\cite{deLeeuw:2017dkd}, however, we expect the expressions reported here to be a much better starting point for further mining of DCFT data.

This paper is organised as follows.
In section~\ref{Kinematics}, we review the kinematics of a 4d ${\cal N}=4$ SCFT with a 1/2-BPS defect of co-dimension one. 
Subsequently, in section~\ref{Nahmpole}, we specialise to the case of ${\cal N}=4$ SYM theory with Nahm pole boundary conditions. 
Our derivation of the manifestly superconformal two-point functions is presented in section~\ref{Twopoint}.
Finally, we finish with an outlook in section~\ref{Outlook}.

\section{Kinematics and boundary superblocks \label{Kinematics}}

Local operators in a 4d $\CN=4$ SCFT belong to unitary irreducible representations of the superconformal algebra $\mathfrak{psu}(2,2|4)$. 
In addition to the conformal algebra in four dimensions, it includes the R-symmetry $\mathfrak{su}(4)_R$ as a bosonic subalgebra.
A co-dimension one 1/2-BPS defect breaks $\mathfrak{psu}(2,2|4)$ to $\osp(4|4)$. 
This is precisely the symmetry of a would-be 3d $\CN=4$ SCFT on the defect, with R-symmetry $\so(4)_r\cong\su(2)_+\oplus\su(2)_-$.
Correlation functions of local operators in this set-up are highly constrained by symmetries, with ref.~\cite{Liendo:2016ymz} presenting the most extensive analysis so far. 
In our discussion we will focus on correlation functions of bulk 1/2-BPS operators. 
Specifically, consider the 4d $\CN=4$ multiplet denoted as $\CB_{[0,p,0]}$ in refs.~\cite{Dolan:2002zh,Liendo:2016ymz}, whose superprimary is a scalar $\CO_{p,0}$ with conformal dimension $\Delta=p$. 
We soak up the R-symmetry indices with a complex polarisation vector $u_I$ obeying $u^2=\delta^{IJ}u_I u_J=0$ (with $I,J$ being $\mathfrak{so}(6)$ indices) such that 
\begin{equation}
\CO_{p,0}(x;u)\equiv u_{I_1}\cdots u_{I_p}\,\CO^{I_1\ldots I_p}(x)\,.
\end{equation}
It is natural to decompose the polarisation vector $u_I$ in a way that reflects the R-symmetry breaking $\mathfrak{so}(6)\to \mathfrak{so}(3)_+ \oplus \mathfrak{so}(3)_-$ by the defect,\footnote{
The following discussion is analogous to the case of a 1/2-BPS Wilson line in $\CN=4$ SYM theory, whose kinematics was discussed in ref.~\cite{Barrat:2021yvp}.
We are very grateful to Aleix Gimenez-Grau for illuminating discussions on this point and for sharing his private notes~\cite{Gimenez-Grau:kinematics} on the kinematics of 1/2-BPS co-dimension one defects with us.}
\begin{align}
\label{eq:u_vec}
u&=(v,w) \,,& \bar{u}&=(v,-w) \,.
\end{align}
Here, $v_i$ and $w_{\bar{\imath}}$ are $\so(3)_+$ and $\so(3)_-$ vectors, respectively, and we defined the $\so(6)$ vector $\bar u$ for later convenience. 
Tensor structures associated with the defect must be constructed out of $v_i$ and $w_{\bar{\imath}}$, which obey the constraint $v^2=-w^2$ since $u^2=0$. 

The one-point function of the superconformal primary of the $\CB_{[0,p,0]}$ multiplet is\begin{equation}
\langle \CO_{p,0}(x;u) \rangle = a_{p,0} \frac{(u\cdot \bar{u})^{p/2}}{(2x_\perp)^p}\,,
\end{equation}
where we have split the spacetime coordinates into directions parallel and orthogonal to the defect, $x=(x_\parallel,x_\perp)$.
The powers of $u$ and $\bar u$ must be integers, otherwise the one-point function vanishes. 
This implies that $p$ must be an even integer for $\CO_{p,0}$ to have a non-zero one-point function.
This selection rule is noted e.g.\ in ref.~\cite{Liendo:2016ymz}.

Defect local operators sit in representations of $\mathfrak{osp}(4|4)$ and obey the same kinematic constraints as local operators in standard conformal field theory. 
For a defect local operator in a representation $(r_+,r_-)$ of the $\so(3)_+\oplus\so(3)_-$ R-symmetry, we can write
\begin{equation}
\label{eq:defect_ops}
\hat{\CO}_{(r_+,r_-)}(x_\parallel;v,w)=v_{i_1}... v_{i_{r_+}}w^{\ib_1}... w^{\ib_{r_-}}\hat{\CO}^{i_1\ldots i_{r_+}}_{\ib_1\ldots \ib_{r_-}}(x_\parallel)\,,
\end{equation}
where all other quantum numbers are suppressed. 
We normalise our operators such that the coefficient of the two-point function is $(v_1\cdot v_2)^{r_+}(w_1\cdot w_2)^{r_-}$ for equal R-charges, and zero otherwise. 
Since irreducible representations of $\SO(3)$ correspond to symmetric traceless tensors, we must further enforce $v^2=w^2=0$. 
Scaling dimension $\delta$ and spin $s$ are subject to constraints from unitarity, and multiplets may obey shortening conditions.  
Of particular interest to us are the 1/2-BPS representations $(B,\pm)_r$ where the notation is that of ref.~\cite{Dolan:2008vc}.
The label $r$ denotes the $\su(2)_\pm$ charge of the scalar superconformal primary, and its conformal dimension $\delta=r$.
A prominent example of such a multiplet is the displacement multiplet for $r=2$, which contains the displacement operator arising from broken translations and the tilt operator for the broken R-symmetry generators.

Bulk and defect local operators in DCFT need not be independent. 
The spacetime dependence of the mixed scalar bulk-defect two-point functions is fixed by conformal symmetry,
\begin{equation}
\label{eq:mixed_2-pt}
\begin{split}
&\langle\CO_{p,0}(x;u_1)\hat{\CO}_{(r_+,r_-)}(y_\parallel;v_2,w_2)\rangle\\
&=\mu_{p,\hat{\CO}} \frac{(v_1\cdot v_2)^{r_+}(w_1\cdot w_2)^{r_-}(u_1\cdot \bar{u}_1)^{(p-r_+-r_-)/2}}{(|x_\parallel-y_\parallel|^2+x_\perp^2)^\delta (2x_\perp)^{p-\delta}}\,.
\end{split}
\end{equation}
The coefficients $\mu_{p,\hat{\CO}}$ are part of the DCFT data. 
Note that the powers of $u$ and $\bar{u}$ must be non-negative integers. 
Otherwise, the correlator vanishes. 
This imposes the selection rule $(p-r_+-r_-)\in2\mathbb{Z}_{\geq0}$. 

Two-point functions of bulk operators in the presence of a conformal co-dimension one defect are functions of a single conformal cross-ratio
\begin{equation}
\label{eq:conf_cross}
\xi
=\frac{(x-y)^2}{4x_\perp y_\perp}\,.
\end{equation}
For a 1/2-BPS defect in a 4d $\CN=4$ SCFT, there are also two independent R-symmetry cross-ratios $\omega_\pm$, which are related to the polarisation vectors of \eq{u_vec} as follows\footnote{
Our variables $\omega_\pm$ correspond precisely to the R-symmetry cross-ratios of ref.~\cite{Liendo:2016ymz}, $\omega_\pm^{\text{\tiny here}}=w_\pm^{\text{\tiny there}}$.
}
\begin{subequations}
\begin{align}
\frac{1}{2}\left(\omega_++\frac{1}{\omega_+}\right) &= \frac{v_1\cdot v_2}{\sqrt{(v_1\cdot v_1)(v_2\cdot v_2)}} \,,\\
- \frac{1}{2}\left(\omega_-+\frac{1}{\omega_-}\right) & =\frac{w_1\cdot w_2}{\sqrt{(w_1\cdot w_1)(w_2\cdot w_2)}}\,.
\end{align}
\end{subequations}
The two-point function of the superconformal primary of the $\CB_{[0,p,0]}$ multiplet can accordingly be expressed as
\begin{equation} \label{eq:2-pt_fn}
\begin{split}
&\langle\CO_{p_1,0}(x;u_1)\CO_{p_2,0}(y;u_2)\rangle \\
&\qquad\quad =\frac{(u_1\cdot \bar{u}_1)^{p_1/2}(u_2\cdot \bar{u}_2)^{p_2/2}}{(2x_\perp)^{p_1} (2y_\perp)^{p_2}}\CF_{p_1,p_2}(\xi,\omega_+,\omega_-)\,, 
\end{split}
\end{equation}
where $\CF_{p_1,p_2}$ is a function of cross-ratios that is not determined by kinematics alone. 
Rather, it can be expanded in terms of bulk or defect conformal blocks, by using the operator product expansion (OPE) between $\CO_{p_1,0}$ and $\CO_{p_2,0}$ or the defect operator expansion (DOE) of $\CO_{p_1,0}$ and $\CO_{p_2,0}$ separately, corresponding to the exchanges of bulk or defect operators, respectively~\cite{McAvity:1995zd}. 
For scalar representations in 4d, the defect conformal blocks are
\begin{align}
\label{eq:fbdy}
\Ffbdy_\delta(\xi)=\xi^{-\delta} {}_2F_1(\delta,\delta-1,2\delta-2;-1/\xi)\,,
\end{align}
where $_2F_1$ denotes the ordinary hypergeometric function.
Note that we adopt the conventions of ref.~\cite{McAvity:1995zd}, which differ from those of ref.~\cite{Liendo:2016ymz}. 
We will not need the expressions for the bulk conformal blocks, which were also found in ref.~\cite{McAvity:1995zd}.
For superconformal defects, all conformal blocks must organise into larger structures called superblocks, which capture an exchanged superconformal representation. 
The coefficient of each conformal block inside a superblock depends on the R-symmetry cross-ratios through a set of functions called R-symmetry blocks. 
Boundary R-symmetry blocks take a particularly simple form as the dependence on the two cross-ratios $\omega_\pm$ factorises. 
In particular, one can define the functions
\begin{equation}
\label{eq:hbdy}
\Fhbdy_r (\omega_\pm) = \frac{r!}{4^r\left(\frac{1}{2}\right)_r} P_r\left(\frac{1}{2} \left(\omega_\pm+\frac{1}{\omega_\pm}\right)\right),
\end{equation}
where $P_\ell$ denotes the $\ell$-th Legendre polynomial. 
The full R-symmetry block then is $\Fhbdy_{r_+} (\omega_+) \Fhbdy_{r_-} (\omega_-)$, where $(r_+,r_-)$ are the R-charges of the exchanged conformal representation. 
The numerical relative coefficients are fixed by the superconformal Ward identities~\cite{Liendo:2016ymz}. 
E.g.\ for the $(B,+)_r$ multiplet, the superblock is\footnote{Our choice of normalisation for $\Ffbdy$ and $\Fhbdy$ leads to additional numerical factors compared with ref.~\cite{Liendo:2016ymz}, e.g.\ in \eq{c_Bp}. For BPS blocks, however, the full superblocks $\FFbdy$ are identical to the ones of ref.~\cite{Liendo:2016ymz}.
}
\begin{equation}
\label{eq:Bp}
\begin{split}
\FFbdy_{(B,+)_r}&=\Fhbdy_r(\omega_+)\Ffbdy_r(\xi)   \\
&\quad+c_1(r)\Fhbdy_{r-1}(\omega_+)\Fhbdy_1(\omega_-)\Ffbdy_{r+1}(\xi) 
\\&\quad+c_2(r)\Fhbdy_{r-2}(\omega_+)\Ffbdy_{r+2}(\xi)\,,
\end{split}
\end{equation}
where 
\begin{align}
\label{eq:c_Bp}
c_1&=\frac{r}{2(1-2r)},&c_2&=\frac{(r-1)^2r(r+1)}{16(2r-3)(2r-1)^2(2r+1)}.
\end{align}

With this normalisation, the function $\CF_{p_1,p_2}$ in \eq{2-pt_fn} admits the following defect channel decomposition
\begin{equation}
\label{eq:defect_ch}
\CF_{p_1,p_2}=a_{p_1,0} \, a_{p_2,0} + \sum_{\chi} \mu_{p_1\hat{\CO}_\chi}\mu_{p_2\hat{\CO}_\chi} \FFbdy_\chi\,,
\end{equation}
where $\chi$ labels the exchanged superconformal representation, and $\hat{\CO}_\chi$ is the associated superconformal primary.

\section{\texorpdfstring{The Nahm pole defect in ${\cal N}=4$ SYM theory}{The Nahm pole defect in N=4 SYM theory}} \label{Nahmpole}

An introduction to this particular system can be found in several places, e.g.\ refs.~\cite{Nagasaki:2012re,Kristjansen:2012tn,
deLeeuw:2017cop, deLeeuw:2019usb,Linardopoulos:2020jck}, so we will be brief.  
Our starting point is  $\CN=4$ SYM theory in 4d with gauge group $\U(N)$. 
We write the bosonic Lagrangian density in the form
\begin{equation}
\frac{2}{\gYM^2}\tr\left( -\frac{1}{4}F_{\mu\nu}F^{\mu\nu} - \frac{1}{2}D_\mu \phi^I D^\mu \phi^I + \frac{1}{4}[\phi^I,\phi^J]^2 \right),
\end{equation}
where $\phi^I$ denotes the six real adjoint scalars of the theory in the vector representation of the $\so(6)\cong \su(4)_R$ R-symmetry.
We will study a particular co-dimension one defect which is closely related to 
the Nahm pole boundary conditions of $\CN=4$ SYM theory formalised in ref.~\cite{Gaiotto:2008sa}. 
Due to the breaking of $\so(6)\to \so(3)_+\oplus \so(3)_-$, it is natural to split the six real scalars into two groups of three, $\phi_i$ and $\phi_{\ib}$, where $i,\ib =1,2,3$. 
Then the BPS locus is determined by Nahm's equation~\cite{Nahm:1979yw} for the three $\phi_i$.
The equation is solved by taking $\phi_{i}^\mathrm{cl}=-\frac{t^{(N)}_{i}}{x_\perp}$, where the $t^{(N)}_i$ are $N$-dimensional $\su(2)$ representation matrices obeying $[t^{(N)}_i,t^{(N)}_j]=\iu\epsilon_{ijk}t^{(N)}_k$.
The remaining fields $\phi_\ib^\mathrm{cl}$ may be taken to be zero.
The superscripts on $\phi_i^\mathrm{cl}$ and $\phi_\ib^\mathrm{cl}$ emphasise that these field profiles satisfy the classical equations of motion (EOM).
Interpreting the classical profile as a vacuum expectation value (VEV), one notices that the classical solution takes the form of a one-point function in a boundary conformal field theory (BCFT) for an operator with $\Delta=1$, i.e.\ the engineering dimension of the field $\phi_i$.

One can construct a closely related co-dimension one defect by gluing together two copies of $\CN=4$ SYM theory with unitary gauge groups whose ranks differ by some integer $k$. 
Concretely, take the groups to be $G_L=\U(N-k)$ on $x_\perp<0$ and $G_R=\U(N)$ 
on $x_\perp>0$. 
Furthermore, take the classical field to have a non-trivial profile
\begin{align}
\label{eq:vevs}
\phi_i^\mathrm{cl} &= -\frac{1}{x_\perp}t_i \oplus 0_{(N-k)\times(N-k)}\,,&  x_\perp&>0\,,
\end{align}
where the $t_i\equiv t^{(k)}_i$ now define an irreducible $\mathfrak{su}(2)$ embedding in $\mathfrak{u}(k)$.
The remaining fields on either side are given trivial VEVs.
Fields valued in the $(N-k)\times (N-k)$ block receive transparent boundary conditions and are identified with their counterparts for $x_\perp<0$.  
Note, in particular, that we do not introduce novel fields living on the boundary.  
For $k=0$ we hence have standard $\CN=4$ SYM theory everywhere in spacetime.\footnote{
A similar co-dimension one DCFT with no jump in the rank of the gauge group has the fields of $\CN=4$ SYM theory interacting with a 3d fundamental hypermultiplet placed on  the defect by hand~\cite{DeWolfe:2001pq}.
Although it is sometimes referred to as having $k=0$, this model is not related to ours, where the only defect fields are restrictions of bulk fields~\cite{Gaiotto:2008sa,Ipsen:2019jne}. }

This gluing construction is very natural from a string theory perspective.
The defect is engineered by ending a different number of D3 branes on a single D5 brane from either side, where the number of D3 branes differs by $k$.
In AdS$_5$/CFT$_4$~\cite{Maldacena:1997re}, this set-up corresponds to a D5 probe brane wrapping AdS$_4\times S^2$ inside AdS$_5\times S^5$ with magnetic $\U(1)$ worldvolume flux through the $S^2$~\cite{Karch:2000gx, DeWolfe:2001pq}.

One may quantise around the background above by splitting the fields $\phi_I = \phi_I^\mathrm{cl} + \tilde{\phi}_I$, where the $\tilde\phi_I$ denote the fluctuating dynamical fields. 
Expanding the action around the classical solution gives rise to position-dependent mass terms and a complicated mixing problem which was solved in ref.~\cite{Buhl-Mortensen:2016jqo}.  
In particular, the $k\times k$ sub-blocks of fields were expanded in terms of fuzzy spherical harmonics~\cite{Hoppe:1982phd,Madore:1991bw}, $\tilde{\phi}_i=\sum_{\ell,m}(\tilde{\phi}_i)_{\ell,m}\hat{Y}_\ell^m$, where $\ell=0,1,\ldots, k-1$ and $m=-\ell,-\ell+1,\ldots, \ell$. 
In analogy to the ordinary spherical harmonics, they obey
\begin{equation}
    L^2\hat{Y}_\ell^m = \ell(\ell+1)\hat{Y}_\ell^m, \qquad L_3\hat{Y}_\ell^m = m\hat{Y}_\ell^m\,,
\end{equation}
where $L_i = [t_i,\bullet]$ and $L^2=\sum_{i} L_i L_i$. 
However, they do not commute and they can be represented as $k\times k$ matrices with components
\begin{equation}
\label{eq:Y_comps}
[\hat{Y}^m_\ell]_{n,n'}=(-1)^{k-n}\sqrt{2\ell+1}  \begin{pmatrix}
        \frac{k-1}{2} & \ell & \frac{k-1}{2} \\
        n-\frac{k+1}{2} & m & \frac{k+1}{2}-n'
    \end{pmatrix},
\end{equation} 
where the parenthesis denotes the Wigner $3j$ symbol. 
With the normalisation as in \eq{Y_comps}, $\tr \hat{Y}_\ell^m (\hat{Y}_{\ell'}^{m'})^\dagger = \delta_{\ell\ell'}\delta_{mm'}$, where we defined $(\hat{Y}_\ell^m)^\dagger=(-1)^m\hat{Y}_\ell^{-m}$. 
One then finds that the fuzzy spherical harmonics obey a fusion rule
\begin{equation}
\label{eq:fusion}
\hat{Y}_{\ell_1}^{m_1}\hat{Y}_{\ell_2}^{m_2} = \sum_{\ell_3=0}^{k-1} \sum_{m_3=-\ell_3}^{\ell_3} F_{\ell_1 m_1 \ell_2 m_2}^{\ell_3 m_3} \hat{Y}_{\ell_3}^{m_3} \,,
\end{equation}
where the fusion coefficients, whose precise form can be found in ref.~\cite{deLeeuw:2017dkd}, involve Wigner $3j$ and $6j$ symbols.

In the basis of fuzzy spherical harmonics the position-dependent mass terms for the dynamical fields in the action give rise to EOM of the form
\begin{equation}
    \left( -\partial_\mu\partial^\mu + \frac{m^2}{x_\perp^2} \right)K^{m^2}(x,y) = \frac{\gYM^2}{2}\delta^{(4)}(x-y)\,,
\end{equation}
where $m^2$ is a dimensionless function of $\ell$. 
Interactions are treated in perturbation theory in powers of $\gYM^2$. 
The Green's function is related to that of a scalar in AdS$_4$, and reads
\begin{align}\label{eq:K_prop}
K^{m^2}(x,y) &= \frac{\gYM^2}{16\pi^2x_\perp y_\perp} \frac{1}{\binom{2\nu+1}{\nu+1/2}} \nonumber\\
&\times \frac{{}_2F_1\left( \nu-\frac{1}{2}, \nu+\frac{1}{2}, 2\nu+1 \,;\, -\xi^{-1} \right)}{(1+\xi)\xi^{\nu+\frac{1}{2}}}\,,
\end{align}
where $\nu=\sqrt{m^2+1/4}$ and $\xi$ is the conformal cross-ratio defined in \eq{conf_cross}. 
The propagator of the scalar modes $(\tilde\phi_i)_{\ell,m}$ can then be written as a linear combination of $K^{m^2}$ for particular $m^2$ of the form $(\ell+n-1)(\ell+n)$, where $n$ is some non-negative integer. 
See ref.~\cite{Buhl-Mortensen:2016jqo} for its precise form.
In particular when $m^2 = \ell(\ell-1)$, the Green's function is related to the boundary conformal block of \eq{fbdy}, i.e.
\begin{equation} \label{eq:K_prop_l}
    K^\ell(x,y) \equiv K^{m^2=\ell(\ell-1)}(x,y) = \frac{\gYM^2}{16\pi^2x_\perp y_\perp} \frac{\Ffbdy_{\ell+1}(\xi)}{\binom{2\ell}{\ell}}\,.
\end{equation}
Note that this definition also holds for $\ell=0,1$, where the mass vanishes and additional care is required when imposing boundary conditions.

\section{Superconformal two-point functions}
\label{Twopoint}

We now compute \eq{2-pt_fn} in perturbation theory for symmetric traceless operators of the form
\begin{equation}
\CW_p(x;u)= u^{I_1}\cdots u^{I_p}\,\tr \phi_{I_1}\cdots \phi_{I_p}\,.
\end{equation}
We use $\CW_p$ to distinguish the single-trace operator from the set of general 1/2-BPS operators with the same quantum numbers, $\CO_{p,0}$, which also contains multi-trace operators. 
As above, the R-symmetry polarisation vector $u^I$ obeys $u^2=0$ which implements the trace-free condition. 
At lowest order the two-point function of two such operators is found by simply inserting the classical fields. 
It trivially factors into a product of one-point functions~\cite{Buhl-Mortensen:2016jqo}. 
Using the residual R-symmetry to evaluate the trace, we arrive at the following R-covariant expression
\begin{equation}
\begin{split}
\label{eq:one-pt-fns}
    \langle\CW_p(x;u)\rangle &= v^{i_1}\dots v^{i_p}\tr\phi_{i_1}^\mathrm{cl}\dots\phi_{i_p}^\mathrm{cl} \\
    &= -2^{p/2+1}\CB_{p+1}\frac{(u\cdot\bar{u})^{p/2}}{(2x_\perp)^p}\delta_{p,\text{even}}\,,
\end{split}
\end{equation}
where we have defined
\begin{equation}
\label{eq:Bernoulli}
    \CB_n = \frac{1}{n} \,B_n\left(\frac{1-k}{2}\right)
\end{equation}
for odd integers $n$ in terms of Bernoulli polynomials $B_n$.

To proceed to the next order we make a single contraction analogously to ref.~\cite{deLeeuw:2017dkd} and replace the remaining fields with their VEVs given in \eq{vevs}. 
The result for the connected part of the two-point function is
\begin{align} 
\label{eq:twopointansatz}
  &\langle\CW_{p_1}(x,u_1)\CW_{p_2}(y,u_2)\rangle_\mathrm{conn.} = \frac{(-1)^{p_1+p_2}\,p_1p_2 \,u_1^Iu_2^J}{(x_\perp)^{p_1-1}(y_\perp)^{p_2-1}} \nonumber \\
& \times  \sum_{\ell,\ell',m,m'} (-1)^{m'} \tr((v_1\cdot t)^{p_1-1}\hat{Y}_\ell^m) \tr((v_2\cdot t)^{p_2-1}\hat{Y}_{\ell'}^{-m'}) \nonumber\\
& \times \,\,\,\,\, \langle(\tilde\phi_I)_{\ell m}(\tilde\phi_J)_{\ell'm'}^\dagger\rangle\,,  
\end{align}
where the sums are finite and over $\ell,\ell'=0,1,\ldots,k-1$, whereas $m=-\ell, -\ell+1,\ldots,\ell$, and similarly for $m'$.
In writing \eq{twopointansatz}, we used the split of R-symmetry polarisation vectors $u^I_{1,2}$ into $v^i_{1,2}$ and $w^\ib_{1,2}$ as in \eq{u_vec}, where the $\so(6)$ vector index $I$ is split into two $\so(3)$ indices $i$ and $\ib$ as discussed above \eq{u_vec}.
Since only $\phi_i$ acquire a VEV, only the $v^i$ part of the polarisation vector is directly contracted into the traces over $\su(2)$ representation matrices $t_i$.
The factor $p_1p_2$ is combinatorial. The propagator of the scalar modes was found in ref.~\cite{Buhl-Mortensen:2016jqo}, and it can be written in terms of the Green's function $K^{m^2}$ in \eq{K_prop}.
The challenge in evaluating \eq{twopointansatz} for arbitrary polarisation vectors $u^I_{1,2}$ comes from computing the traces left over by the VEVs and the fuzzy spherical harmonic.
In principle, the traces can be computed for any choice of $v_{1,2}$ using the fusion rule~\eqref{eq:fusion}. 
However, it quickly becomes computationally expensive and a closed form expression is not immediate. 
Choosing the normalisation $v^2=1$, we observe by explicit computation that 
\begin{equation}
\label{eq:traces}
	\tr((v\cdot t)^L\hat{Y}_\ell^m) = \alpha^L_\ell \sqrt{\frac{(\ell-m)!}{(\ell+m)!}} \left( \frac{v^1+\iu v^2}{\sqrt{1-\left(v^3\right)^2}} \right)^m P_\ell^m(v^3)
\end{equation}
for non-negative $m$.  
For negative $m$ we find that one must first perform the substitution $\frac{\sqrt{1-(v^3)^2}}{v^1+\iu v^2} = \frac{v^1-\iu v^2}{\sqrt{1-(v^3)^2}}$ as if $v$ was real, and then proceed as before.
The functions $P_\ell^m$ are associated Legendre polynomials and the coefficients $\alpha_\ell^L$ were defined in ref.~\cite{deLeeuw:2017dkd} by the expansion 
\begin{equation}
\label{eq:def_alpha}
	t_3^L = \sum_{\ell=0}^L \alpha_\ell^L\hat{Y}_\ell^0.
\end{equation}
We have checked \eq{traces} for $L\leq 6$, after which applying the fusion rule \eqref{eq:fusion} becomes computationally expensive. It would be interesting to prove the formula.

In ref.~\cite{deLeeuw:2017dkd}, two separate closed form expressions for the coefficients $\alpha_\ell^L$ were given.
The first one involves a sum of products of the fusion coefficients in \eq{fusion} while the second one is obtained via a recursion relation.
Even though the second expression is of lower complexity, it still involves the inversion of an $\lfloor\ell/2\rfloor\times \lfloor\ell/2\rfloor$ matrix whose entries are Bernoulli polynomials $\CB_n$ where $n\leq\ell$ is an odd integer.
Here we report a much simpler expression for the coefficients based on the simple observation that $\alpha_\ell^L=\tr(t_3^L\hat{Y}_\ell^0)$, which follows from \eq{def_alpha} and the orthonormality of the fuzzy spherical harmonics reported below \eq{Y_comps}.
Using the explicit form of $\hat{Y}^0_\ell$ in \eq{Y_comps}, one finds the following expression for $\alpha_\ell^L$ in terms of a simple sum over Wigner $3j$ symbols
\begin{equation}
\begin{split}
	\alpha_\ell^L 
	&= (-1)^k\sqrt{2\ell+1} \sum_{n=1}^k  (-1)^n\left( \frac{k+1}{2} - n \right)^{L}   \\
&  \quad \times  \begin{pmatrix}
        \frac{k-1}{2} & \ell & \frac{k-1}{2} \\
        n-\frac{k+1}{2} & 0 & \frac{k+1}{2}-n
    \end{pmatrix}.
    \end{split}
\end{equation}
An important property of these coefficients is that they vanish for $\ell>L$, as well as when $L+\ell$ is odd.
Additionally they satisfy the completeness relation 
\begin{equation}
\label{eq:alpha_comp}
    \sum_{\ell=0}^{\min(L_1,L_2)} \alpha_\ell^{L_1}\alpha_\ell^{L_2} = -2\CB_{L_1+L_2+1}\,,
\end{equation}
which was first noted in ref.~\cite{deLeeuw:2017dkd}. 

Plugging the expressions for the propagators into \eq{twopointansatz} we obtain
\begin{align}
    &\langle \CW_{p_1}(x;u_1)\CW_{p_2}(y;u_2) \rangle_\mathrm{conn.} = \frac{p_1p_2}{(x_\perp)^{p_1-1}(y_\perp)^{p_2-1}} \nonumber\\
    &\times\sum_{\ell=0}^{p_{\text{\tiny min}}-1} \biggl[ (v_1\cdot v_2) A_\ell^{(1)} 
    \left( \frac{\ell+1}{2\ell+1}K^\ell + \frac{\ell}{2\ell+1}K^{\ell+2} \right) \\
    &\qquad+A_\ell^{(2)} \frac{1}{2\ell+1}\left( K^{\ell} - K^{\ell+2} \right) + (w_1\cdot w_2)A_\ell^{(1)}K^{\ell+1} \biggr], \nonumber
\end{align}
where
\begin{subequations}
\begin{align}
A_\ell^{(1)} &= \phantom{-}\sum_{m=-\ell}^\ell (-1)^m \nonumber \\
& \qquad \times\tr((v_1\cdot t)^{p_1-1}\hat{Y}_\ell^m)\tr((v_2\cdot t)^{p_2-1}\hat{Y}_\ell^{-m}) \\
A_\ell^{(2)} &= -\iu\sum_{m,m'} (-1)^{m'} \nonumber\\ 
&\qquad \times\tr((v_1\cdot t)^{p_1-1}\hat{Y}_\ell^m)\tr((v_2\cdot t)^{p_2-1}\hat{Y}_\ell^{-m'}) \nonumber\\
    &\qquad \times\varepsilon_{ijk}v_1^iv_2^j[t_k^{(2\ell+1)}]_{\ell-m+1,\ell-m'+1}\,,
\end{align}
\end{subequations}
and we defined $p_{\text{\tiny min}}=\min(p_1,p_2)$ for brevity.
Here, $\varepsilon_{ijk}$ denotes the totally anti-symmetric tensor with $\varepsilon_{123}=+1$, and $[t_k^{(2\ell+1)}]_{n,n'}$ are the components of the $\su(2)$ generators in the $(2\ell+1)$-dimensional representation.
In order to perform the sums it is convenient to use the residual R-symmetry to set $v_1=(0,0,1)$ and to identify $v_2^3 \equiv v_1\cdot v_2$ at the end.
In doing so, the sums over $m$ trivialise, since all but the $m=0$ terms vanish, and we can easily read off 
\begin{equation}
	A_\ell^{(1)} = \alpha_\ell^{p_1-1}\alpha_\ell^{p_2-1}P_\ell(v_1\cdot v_2)\,,
\end{equation}
where $P_\ell$ is the $\ell$-th Legendre polynomial.
Similarly, a short calculation involving the associated Legendre polynomial identity $P_\ell^{-1} = -\frac{(\ell-1)!}{(\ell+1)!}P_\ell^1$ shows
\begin{equation}
\begin{split}
	A_\ell^{(2)} &= \alpha_\ell^{p_1-1}\alpha_\ell^{p_2-1}\sqrt{1-(v_1\cdot v_2)^2}P^1_\ell(v_1\cdot v_2) \\
	&= \alpha_\ell^{p_1-1}\alpha_\ell^{p_2-1} \left( (v_1\cdot v_2)^2 - 1 \right)  P'_\ell(v_1\cdot v_2)\,,
	\end{split}
\end{equation}
where $P'_\ell(x)\equiv \frac{\dd}{\dd x}P_\ell(x)$.
Finally plugging this back into the correlator, we can simplify the result using the two recursion relations
\begin{subequations}
\begin{align}
    (x^2-1)P'_\ell(x) &= \ell xP_\ell(x) - \ell P_{\ell-1}(x), \\
    (\ell+1)P_{\ell+1}(x) &= (2\ell+1)xP_\ell(x) - \ell P_{\ell-1}(x)
\end{align}
\end{subequations}
to obtain
\begin{align}
\label{eq:cov_2-pt-fn}
    &\langle \CW_{p_1}(x;u_1)\CW_{p_2}(y;u_2) \rangle_\mathrm{conn.} = \frac{p_1p_2}{(x_\perp)^{p_1-1}(y_\perp)^{p_2-1}} \nonumber\\
    &\times\sum_{\ell=0}^{p_{\text{\tiny min}}-1} \alpha_\ell^{p_1-1}\alpha_\ell^{p_2-1} \Biggl[P_\ell(v_1\cdot v_2)P_1(w_1\cdot w_2) K^{\ell+1}  \\
    &+ \frac{\ell+1}{2\ell+1}P_{\ell+1}(v_1\cdot v_2) K^\ell + \frac{\ell}{2\ell+1} P_{\ell-1}(v_1\cdot v_2) K^{\ell+2} \Biggr]\,. \nonumber
\end{align}
Note that the correlator vanishes when $p_1+p_2$ is odd, so we have dropped a factor of $(-1)^{p_1+p_2}$.

Since \eq{cov_2-pt-fn} is manifestly R-symmetry covariant, one must be able to decompose it into superconformal blocks. 
By virtue of the relation \eqref{eq:K_prop_l} between the Green's function and the boundary conformal blocks as well as the form of the boundary R-symmetry blocks built out of \eq{hbdy}, \eq{cov_2-pt-fn} can be recast in terms of boundary superblocks.
Accounting for the relative factors between different products of spacetime and R-symmetry blocks as in \eq{Bp}, one finds the very pleasing expression
\begin{equation}
\label{eq:superblock_2-pt-fn}
\begin{split}
    &\langle\CW_{p_1}(x;u_1)\CW_{p_2}(y;u_2)\rangle_\mathrm{conn.} \\
    &= \frac{(u_1\cdot \bar{u}_1)^{p_1/2}(u_2\cdot \bar{u}_2)^{p_2/2}}{(2x_\perp)^{p_1}(2y_\perp)^{p_2}}\,\frac{p_1p_22^{(p_1+p_2)/2}\gYM^2}{8\pi^2}\,  \\
  & \quad  \times\sum_{\ell=0}^{p_{\text{\tiny min}}-1} \alpha_\ell^{p_1-1}\alpha_\ell^{p_2-1} \mathfrak{F}^\mathrm{bdy}_{(B,+)_{\ell+1}}\,,
\end{split}
\end{equation}
where $\mathfrak{F}^\mathrm{bdy}_\chi$ denotes the superblock for the boundary superconformal representation $\chi$.
This is our final result, expressing the two-point function at finite $N$ in a manifestly superconformal language.
Comparing with eqs.~\eqref{eq:2-pt_fn} and \eqref{eq:defect_ch}, one can read off all the leading-order bulk-to-defect couplings $\mu_{p_{1,2},\hat{\CO}}$ up to degeneracies.

We emphasise that this sum over boundary superblocks is not only finite, but it also contains only 1/2-BPS blocks of the $(B,+)$ type introduced below \eq{defect_ops}. 
Note that this is a subset of the blocks allowed by the R-symmetry selection rule introduced below \eq{mixed_2-pt}, which generally allows for 1/4-BPS and long blocks~\cite{Liendo:2016ymz}.
The appearance of $(B,+)$ type multiplets only at this order in perturbation theory could be anticipated by the following argument.
The R-symmetry block corresponding to an operator in the $(r_+,r_-)$ representation of $\su(2)_+\oplus\su(2)_-$, $\Fhbdy_{r_+}\Fhbdy_{r_-}\sim P_{r_+}(v_1\cdot v_2)P_{r_-}(w_1\cdot w_2)$, involves a term of order $(w_1\cdot w_2)^{r_-}$.
However such a term can only be obtained by making at least $r_-$ contractions, since the $\phi_\ib$ have vanishing VEVs.
At order $\gYM^2$ we can therefore exclude any superblocks which contain R-symmetry representations with $r_->1$.
Absent any cancellations between longer blocks, this leaves just the $(B,+)$ blocks, as well as the $(B,-)_1$ block, which we have now also ruled out.
More generally the statement is that at order $\gYM^{2n}$, only R-symmetry representations $(r_+,r_-)$ with $r_-\leq n$ are allowed, which gives a constraint on the highest possible $\su(2)_-$ spin of the superprimaries.
For 1/2-BPS blocks we must have $r_-\leq n$, for 1/4-BPS blocks $r_-<n$, and for long blocks $r_-<n-1$.
This constraint becomes redundant once $n=p_{\text{\tiny min}}$, at which point the selection rules given in ref.~\cite{Liendo:2016ymz} also exclude higher weight superblocks.

As a special case, consider the two-point function of the superprimary of the stress tensor multiplet, i.e.\ $p_1=p_2=2$ in \eq{superblock_2-pt-fn}. 
The sum on the right-hand side truncates and involves only a single boundary superblock $(B,+)_2$, which accommodates the displacement operator $D(x_\parallel)$ and tilt operators $t_a(x_\parallel)$ for the broken R-symmetry generators.
The normalisation of the stress tensor and displacement multiplets are physically meaningful, with their bulk-to-boundary couplings equal to 1.
Consequently, the coefficient of $\mathfrak{F}^\mathrm{bdy}_{(B,+)_{2}}$ in \eq{superblock_2-pt-fn} is proportional to the two-point function of the displacement multiplet.
The $k$-dependence of the two-point function of the displacement and tilt operators, $c_D$ and $c_t$, respectively, are then determined to be $c_D\propto c_t\propto -\gYM^2 \CB_3$ at leading order in $\gYM^2$, where we used the completeness relation~\eqref{eq:alpha_comp}.
The proportionality factors are positive $\CO(1)$ numbers fixed by the superconformal Ward identities, which we shall not determine here. 
The coefficients $c_D$ and $c_t$ are positive for integer $k\geq2$, which is consistent with unitarity.
They determine a boundary Weyl anomaly coefficient~\cite{Herzog:2015ioa,Herzog:2017xha,Herzog:2017kkj}\footnote{
See also ref.~\cite{Chalabi:2021jud} (and refs.\ therein) for defect Weyl anomalies across dimensions and their relations to DCFT correlation functions.
}
and the metric on the defect conformal manifold $\SU(4)/(\SU(2)_+\times\SU(2)_-)$~\cite{Drukker:2022pxk,Herzog:2023dop}.
Our results appear consistent with ref.~\cite{deLeeuw:2023wjq}, which computed the full stress tensor two-point function in the presence of the defect and extracted $c_D$ from it directly.

The extension of \eq{superblock_2-pt-fn} to multi-trace operators is straightforward. 
Making a single contraction between two trace factors amounts to inserting the two-point function \eq{superblock_2-pt-fn} of the corresponding single-trace operators.
The remaining factors are then simply evaluated as one-point functions to lowest order (see \eq{one-pt-fns}), and one sums over all possible combinations.

\section{Outlook\label{Outlook}}

In this letter, we derived a manifestly superconformal expression for the leading-order two-point function of all chiral primary operators.
See \eq{superblock_2-pt-fn} for our main result. 
Our perturbative computation naturally organises in a defect channel expansion of superblocks, which allowed us to extract the leading-order bulk-to-defect couplings of all operators in the DOE of single-trace chiral primaries $\CW_p$.
This sets the scene for scrutinising the possibility of solving the Nahm pole via the defect conformal bootstrap programme. 

Given our complete knowledge of one side of the crossing equation at leading order, this would be the ideal set-up for an inversion formula which could extract all bulk data at finite $N$.
While this is possible for defects of co-dimension two or higher, see e.g.\ refs.~\cite{Caron-Huot:2017vep,Lemos:2017vnx, Isachenkov:2018pef,Liendo:2019jpu}, these techniques are not immediately applicable to our model which has co-dimension one. 
We emphasise that although it is often believed to be simpler, the case of co-dimension one poses conceptual challenges that are yet to be fully addressed from an analytic bootstrap perspective. See refs.~\cite{Mazac:2018biw, Bianchi:2022ppi} for the current state-of-the-art.
Absent such techniques, one could attempt a generalisation of the approach of ref.~\cite{Bissi:2018mcq} to the Wilson-Fisher BCFT, which was successfully applied to a simple superconformal system in ref.~\cite{Gimenez-Grau:2020jvf}.\footnote{
We are grateful to Philine van Vliet for bringing this work to our attention.} 
Unlike the case there, however, preliminary analysis suggests that there are infinitely many contributions that enter the bulk side of the crossing equation at the same order in $\gYM^2$. 
This poses an additional obstacle to applying this method. 
Constraints could also be obtained by using the fact that only finitely many operators appear in the DOE of $\CW_p$ at leading order. 
In simple DCFTs with finite DOEs, refs.~\cite{Lauria:2020emq,Behan:2020nsf,Behan:2021tcn,Behan:2023ile} were able to derive exact OPE relations by exploiting analyticity properties of mixed bulk-defect correlation functions. 
It would be interesting to investigate if similar constraints can be derived for the present model.\footnote{
We are grateful to Edoardo Lauria for raising this question.}
Alternatively, one could make progress by focussing on a protected subsector. 
In particular, the sector of 1/2-BPS bulk and defect operators has an algebraic crossing equation with finitely many terms~\cite{Liendo:2016ymz}, which one may hope to solve.
Moreover, this system possesses a free parameter $k$, which denotes the difference between the ranks of the gauge group on either side of the defect. 
This makes it particularly attractive for finite-$N$ bootstrap, as the additional $k$-dependence of the DCFT data could resolve degeneracies as in ref.~\cite{deLeeuw:2017dkd}.

Another possibility would be to scrutinise the crossing equations at large $N$.
Concretely, one could envision obtaining higher-order bulk-to-defect couplings using large-$N$ conformal data as input. 
These include one-point functions and anomalous dimensions, which are known to any loop order~\cite{Minahan:2002ve,Buhl-Mortensen:2017ind,Komatsu:2020sup,Gombor:2020kgu,Kristjansen:2020mhn,
Gombor:2020auk}, and structure constants, which are likewise (at least in principle) calculable to any loop order~\cite{Basso:2015zoa,Komatsu:2017buu} from integrability.  
In the large-$N$ limit certain correlation functions can be calculated in the dual string theory picture as well, albeit in a semi-classical approximation~\cite{Nagasaki:2012re,Georgiou:2023yak}. 
Furthermore, it may be possible to apply the holographic bootstrap method of ref.~\cite{Gimenez-Grau:2023fcy} to this system.

Along the lines of the present work it would be an interesting challenge to derive the manifestly superconformal expression for the complete set of two-point functions of chiral primary operators in 3d ABJM theory~\cite{Aharony:2008ug} with a Nahm pole defect, which was studied from the point of view of integrability in~\cite{Minahan:2008hf,Kristjansen:2021abc,Gombor:2022aqj}.
In this case the bulk and defect R-symmetry conformal blocks are not known and would have to be derived in the spirit of ref.~\cite{Liendo:2016ymz}.

\vspace{0.3cm}

\section*{Acknowledgments}
We would like to thank Martin Ammon, Julien Barrat, Agnese Bissi, Aleix Gimenez-Grau, Jakob Hollweck, Shota Komatsu, Edoardo Lauria, Pedro Liendo, Volker Schomerus, Charlotte Sleight, Jake Stedman, Maxime Tr\'epanier, Cristian Vergu, Philine van Vliet, Matthias Wilhelm, and Kostya Zarembo for valuable discussions. 
We are particularly grateful to Julien Barrat for sharing his thesis, to Agnese Bissi for useful comments on a preliminary version of the manuscript, and to Aleix Gimenez-Grau for sharing his private notes with us. 
A.C.\ and C.K.\ were supported by DFF-FNU through grant number 1026-00103B.

\bibliographystyle{elsarticle-num} 

\bibliography{two_pt_microboot_bib}

\begin{thebibliography}{10}
\expandafter\ifx\csname url\endcsname\relax
  \def\url#1{\texttt{#1}}\fi
\expandafter\ifx\csname urlprefix\endcsname\relax\def\urlprefix{URL }\fi
\expandafter\ifx\csname href\endcsname\relax
  \def\href#1#2{#2} \def\path#1{#1}\fi

\bibitem{deLeeuw:2015hxa}
M.~de~Leeuw, C.~Kristjansen, K.~Zarembo, {One-point Functions in Defect CFT and
  Integrability}, JHEP 08 (2015) 098.
\newblock \href {http://arxiv.org/abs/1506.06958} {\path{arXiv:1506.06958}},
  \href {https://doi.org/10.1007/JHEP08(2015)098}
  {\path{doi:10.1007/JHEP08(2015)098}}.

\bibitem{Buhl-Mortensen:2015gfd}
I.~Buhl-Mortensen, M.~de~Leeuw, C.~Kristjansen, K.~Zarembo, {One-point
  Functions in AdS/dCFT from Matrix Product States}, JHEP 02 (2016) 052.
\newblock \href {http://arxiv.org/abs/1512.02532} {\path{arXiv:1512.02532}},
  \href {https://doi.org/10.1007/JHEP02(2016)052}
  {\path{doi:10.1007/JHEP02(2016)052}}.

\bibitem{Buhl-Mortensen:2017ind}
I.~Buhl-Mortensen, M.~de~Leeuw, A.~C. Ipsen, C.~Kristjansen, M.~Wilhelm,
  {Asymptotic One-Point Functions in Gauge-String Duality with Defects}, Phys.
  Rev. Lett. 119~(26) (2017) 261604.
\newblock \href {http://arxiv.org/abs/1704.07386} {\path{arXiv:1704.07386}},
  \href {https://doi.org/10.1103/PhysRevLett.119.261604}
  {\path{doi:10.1103/PhysRevLett.119.261604}}.

\bibitem{Gombor:2020kgu}
T.~Gombor, Z.~Bajnok, {Boundary states, overlaps, nesting and bootstrapping
  AdS/dCFT}, JHEP 10 (2020) 123.
\newblock \href {http://arxiv.org/abs/2004.11329} {\path{arXiv:2004.11329}},
  \href {https://doi.org/10.1007/JHEP10(2020)123}
  {\path{doi:10.1007/JHEP10(2020)123}}.

\bibitem{Komatsu:2020sup}
S.~Komatsu, Y.~Wang, {Non-perturbative defect one-point functions in planar
  $\mathcal{N}=4$ super-Yang-Mills}, Nucl. Phys. B 958 (2020) 115120.
\newblock \href {http://arxiv.org/abs/2004.09514} {\path{arXiv:2004.09514}},
  \href {https://doi.org/10.1016/j.nuclphysb.2020.115120}
  {\path{doi:10.1016/j.nuclphysb.2020.115120}}.

\bibitem{Kristjansen:2020mhn}
C.~Kristjansen, D.~M\"uller, K.~Zarembo, {Integrable boundary states in D3-D5
  dCFT: beyond scalars}, JHEP 08 (2020) 103.
\newblock \href {http://arxiv.org/abs/2005.01392} {\path{arXiv:2005.01392}},
  \href {https://doi.org/10.1007/JHEP08(2020)103}
  {\path{doi:10.1007/JHEP08(2020)103}}.

\bibitem{Gombor:2020auk}
T.~Gombor, Z.~Bajnok, {Boundary state bootstrap and asymptotic overlaps in
  AdS/dCFT}, JHEP 03 (2021) 222.
\newblock \href {http://arxiv.org/abs/2006.16151} {\path{arXiv:2006.16151}},
  \href {https://doi.org/10.1007/JHEP03(2021)222}
  {\path{doi:10.1007/JHEP03(2021)222}}.

\bibitem{Wang:2020seq}
Y.~Wang, {Taming defects in $ \mathcal{N} $ = 4 super-Yang-Mills}, JHEP 08~(08)
  (2020) 021.
\newblock \href {http://arxiv.org/abs/2003.11016} {\path{arXiv:2003.11016}},
  \href {https://doi.org/10.1007/JHEP08(2020)021}
  {\path{doi:10.1007/JHEP08(2020)021}}.

\bibitem{McAvity:1995zd}
D.~M. McAvity, H.~Osborn, {Conformal field theories near a boundary in general
  dimensions}, Nucl. Phys. B 455 (1995) 522--576.
\newblock \href {http://arxiv.org/abs/cond-mat/9505127}
  {\path{arXiv:cond-mat/9505127}}, \href
  {https://doi.org/10.1016/0550-3213(95)00476-9}
  {\path{doi:10.1016/0550-3213(95)00476-9}}.

\bibitem{Liendo:2012hy}
P.~Liendo, L.~Rastelli, B.~C. van Rees, {The Bootstrap Program for Boundary
  CFT$_d$}, JHEP 07 (2013) 113.
\newblock \href {http://arxiv.org/abs/1210.4258} {\path{arXiv:1210.4258}},
  \href {https://doi.org/10.1007/JHEP07(2013)113}
  {\path{doi:10.1007/JHEP07(2013)113}}.

\bibitem{Billo:2016cpy}
M.~Bill\`o, V.~Gon\c{c}alves, E.~Lauria, M.~Meineri, {Defects in conformal
  field theory}, JHEP 04 (2016) 091.
\newblock \href {http://arxiv.org/abs/1601.02883} {\path{arXiv:1601.02883}},
  \href {https://doi.org/10.1007/JHEP04(2016)091}
  {\path{doi:10.1007/JHEP04(2016)091}}.

\bibitem{deLeeuw:2017dkd}
M.~de~Leeuw, A.~C. Ipsen, C.~Kristjansen, K.~E. Vardinghus, M.~Wilhelm,
  {Two-point functions in AdS/dCFT and the boundary conformal bootstrap
  equations}, JHEP 08 (2017) 020.
\newblock \href {http://arxiv.org/abs/1705.03898} {\path{arXiv:1705.03898}},
  \href {https://doi.org/10.1007/JHEP08(2017)020}
  {\path{doi:10.1007/JHEP08(2017)020}}.

\bibitem{Caron-Huot:2017vep}
S.~Caron-Huot, {Analyticity in Spin in Conformal Theories}, JHEP 09 (2017) 078.
\newblock \href {http://arxiv.org/abs/1703.00278} {\path{arXiv:1703.00278}},
  \href {https://doi.org/10.1007/JHEP09(2017)078}
  {\path{doi:10.1007/JHEP09(2017)078}}.

\bibitem{Lemos:2017vnx}
M.~Lemos, P.~Liendo, M.~Meineri, S.~Sarkar, {Universality at large transverse
  spin in defect CFT}, JHEP 09 (2018) 091.
\newblock \href {http://arxiv.org/abs/1712.08185} {\path{arXiv:1712.08185}},
  \href {https://doi.org/10.1007/JHEP09(2018)091}
  {\path{doi:10.1007/JHEP09(2018)091}}.

\bibitem{Isachenkov:2018pef}
M.~Isachenkov, P.~Liendo, Y.~Linke, V.~Schomerus, {Calogero-Sutherland Approach
  to Defect Blocks}, JHEP 10 (2018) 204.
\newblock \href {http://arxiv.org/abs/1806.09703} {\path{arXiv:1806.09703}},
  \href {https://doi.org/10.1007/JHEP10(2018)204}
  {\path{doi:10.1007/JHEP10(2018)204}}.

\bibitem{Liendo:2019jpu}
P.~Liendo, Y.~Linke, V.~Schomerus, {A Lorentzian inversion formula for defect
  CFT}, JHEP 08 (2020) 163.
\newblock \href {http://arxiv.org/abs/1903.05222} {\path{arXiv:1903.05222}},
  \href {https://doi.org/10.1007/JHEP08(2020)163}
  {\path{doi:10.1007/JHEP08(2020)163}}.

\bibitem{Behan:2023ile}
C.~Behan, E.~Lauria, M.~Nocchi, P.~van Vliet, {Analytic and numerical bootstrap
  for the long-range Ising model} (11 2023).
\newblock \href {http://arxiv.org/abs/2311.02742} {\path{arXiv:2311.02742}}.

\bibitem{Bertucci:2022ptt}
F.~Bertucci, J.~Henriksson, B.~McPeak, {Analytic bootstrap of mixed correlators
  in the O(n) CFT}, JHEP 10 (2022) 104.
\newblock \href {http://arxiv.org/abs/2205.09132} {\path{arXiv:2205.09132}},
  \href {https://doi.org/10.1007/JHEP10(2022)104}
  {\path{doi:10.1007/JHEP10(2022)104}}.

\bibitem{Gimenez-Grau:2022ebb}
A.~Gimenez-Grau, {Probing magnetic line defects with two-point functions} (12
  2022).
\newblock \href {http://arxiv.org/abs/2212.02520} {\path{arXiv:2212.02520}}.

\bibitem{Bianchi:2022sbz}
L.~Bianchi, D.~Bonomi, E.~de~Sabbata, {Analytic bootstrap for the localized
  magnetic field}, JHEP 04 (2023) 069.
\newblock \href {http://arxiv.org/abs/2212.02524} {\path{arXiv:2212.02524}},
  \href {https://doi.org/10.1007/JHEP04(2023)069}
  {\path{doi:10.1007/JHEP04(2023)069}}.

\bibitem{Bianchi:2023gkk}
L.~Bianchi, D.~Bonomi, E.~de~Sabbata, A.~Gimenez-Grau, {Analytic bootstrap for
  magnetic impurities} (12 2023).
\newblock \href {http://arxiv.org/abs/2312.05221} {\path{arXiv:2312.05221}}.

\bibitem{Gimenez-Grau:2019hez}
A.~Gimenez-Grau, P.~Liendo, {Bootstrapping line defects in $\mathcal{N}=2$
  theories}, JHEP 03 (2020) 121.
\newblock \href {http://arxiv.org/abs/1907.04345} {\path{arXiv:1907.04345}},
  \href {https://doi.org/10.1007/JHEP03(2020)121}
  {\path{doi:10.1007/JHEP03(2020)121}}.

\bibitem{Barrat:2020vch}
J.~Barrat, P.~Liendo, J.~Plefka, {Two-point correlator of chiral primary
  operators with a Wilson line defect in $ \mathcal{N} $ = 4 SYM}, JHEP 05
  (2021) 195.
\newblock \href {http://arxiv.org/abs/2011.04678} {\path{arXiv:2011.04678}},
  \href {https://doi.org/10.1007/JHEP05(2021)195}
  {\path{doi:10.1007/JHEP05(2021)195}}.

\bibitem{Barrat:2021yvp}
J.~Barrat, A.~Gimenez-Grau, P.~Liendo, {Bootstrapping holographic defect
  correlators in $ \mathcal{N} $ = 4 super Yang-Mills}, JHEP 04 (2022) 093.
\newblock \href {http://arxiv.org/abs/2108.13432} {\path{arXiv:2108.13432}},
  \href {https://doi.org/10.1007/JHEP04(2022)093}
  {\path{doi:10.1007/JHEP04(2022)093}}.

\bibitem{Mazac:2018biw}
D.~Maz\'a\v{c}, L.~Rastelli, X.~Zhou, {An analytic approach to BCFT$_{d}$},
  JHEP 12 (2019) 004.
\newblock \href {http://arxiv.org/abs/1812.09314} {\path{arXiv:1812.09314}},
  \href {https://doi.org/10.1007/JHEP12(2019)004}
  {\path{doi:10.1007/JHEP12(2019)004}}.

\bibitem{Liendo:2016ymz}
P.~Liendo, C.~Meneghelli, {Bootstrap equations for $ \mathcal{N} $ = 4 SYM with
  defects}, JHEP 01 (2017) 122.
\newblock \href {http://arxiv.org/abs/1608.05126} {\path{arXiv:1608.05126}},
  \href {https://doi.org/10.1007/JHEP01(2017)122}
  {\path{doi:10.1007/JHEP01(2017)122}}.

\bibitem{Dolan:2002zh}
F.~A. Dolan, H.~Osborn, {On short and semi-short representations for
  four-dimensional superconformal symmetry}, Annals Phys. 307 (2003) 41--89.
\newblock \href {http://arxiv.org/abs/hep-th/0209056}
  {\path{arXiv:hep-th/0209056}}, \href
  {https://doi.org/10.1016/S0003-4916(03)00074-5}
  {\path{doi:10.1016/S0003-4916(03)00074-5}}.

\bibitem{Gimenez-Grau:kinematics}
A.~Gimenez-Grau, Private communications (2023).

\bibitem{Dolan:2008vc}
F.~A. Dolan, {On Superconformal Characters and Partition Functions in Three
  Dimensions}, J. Math. Phys. 51 (2010) 022301.
\newblock \href {http://arxiv.org/abs/0811.2740} {\path{arXiv:0811.2740}},
  \href {https://doi.org/10.1063/1.3211091} {\path{doi:10.1063/1.3211091}}.

\bibitem{Nagasaki:2012re}
K.~Nagasaki, S.~Yamaguchi, {Expectation values of chiral primary operators in
  holographic interface CFT}, Phys. Rev. D 86 (2012) 086004.
\newblock \href {http://arxiv.org/abs/1205.1674} {\path{arXiv:1205.1674}},
  \href {https://doi.org/10.1103/PhysRevD.86.086004}
  {\path{doi:10.1103/PhysRevD.86.086004}}.

\bibitem{Kristjansen:2012tn}
C.~Kristjansen, G.~W. Semenoff, D.~Young, {Chiral primary one-point functions
  in the D3-D7 defect conformal field theory}, JHEP 01 (2013) 117.
\newblock \href {http://arxiv.org/abs/1210.7015} {\path{arXiv:1210.7015}},
  \href {https://doi.org/10.1007/JHEP01(2013)117}
  {\path{doi:10.1007/JHEP01(2013)117}}.

\bibitem{deLeeuw:2017cop}
M.~de~Leeuw, A.~C. Ipsen, C.~Kristjansen, M.~Wilhelm, {Introduction to
  integrability and one-point functions in $\mathcal N=$ 4 supersymmetric
  Yang\textendash{}Mills theory and its defect cousin} (8 2017).
\newblock \href {http://arxiv.org/abs/1708.02525} {\path{arXiv:1708.02525}},
  \href {https://doi.org/10.1093/oso/9780198828150.003.0008}
  {\path{doi:10.1093/oso/9780198828150.003.0008}}.

\bibitem{deLeeuw:2019usb}
M.~de~Leeuw, {One-point functions in AdS/dCFT}, J. Phys. A 53~(28) (2020)
  283001.
\newblock \href {http://arxiv.org/abs/1908.03444} {\path{arXiv:1908.03444}},
  \href {https://doi.org/10.1088/1751-8121/ab15fb}
  {\path{doi:10.1088/1751-8121/ab15fb}}.

\bibitem{Linardopoulos:2020jck}
G.~Linardopoulos, {Solving holographic defects}, PoS CORFU2019 (2020) 141.
\newblock \href {http://arxiv.org/abs/2005.02117} {\path{arXiv:2005.02117}},
  \href {https://doi.org/10.22323/1.376.0141} {\path{doi:10.22323/1.376.0141}}.

\bibitem{Gaiotto:2008sa}
D.~Gaiotto, E.~Witten, {Supersymmetric Boundary Conditions in N=4 Super
  Yang-Mills Theory}, J. Statist. Phys. 135 (2009) 789--855.
\newblock \href {http://arxiv.org/abs/0804.2902} {\path{arXiv:0804.2902}},
  \href {https://doi.org/10.1007/s10955-009-9687-3}
  {\path{doi:10.1007/s10955-009-9687-3}}.

\bibitem{Nahm:1979yw}
W.~Nahm, {A Simple Formalism for the BPS Monopole}, Phys. Lett. B 90 (1980)
  413--414.
\newblock \href {https://doi.org/10.1016/0370-2693(80)90961-2}
  {\path{doi:10.1016/0370-2693(80)90961-2}}.

\bibitem{DeWolfe:2001pq}
O.~DeWolfe, D.~Z. Freedman, H.~Ooguri, {Holography and defect conformal field
  theories}, Phys. Rev. D 66 (2002) 025009.
\newblock \href {http://arxiv.org/abs/hep-th/0111135}
  {\path{arXiv:hep-th/0111135}}, \href
  {https://doi.org/10.1103/PhysRevD.66.025009}
  {\path{doi:10.1103/PhysRevD.66.025009}}.

\bibitem{Ipsen:2019jne}
A.~C. Ipsen, K.~E. Vardinghus, {The dilatation operator for defect conformal
  $N=4$ SYM} (9 2019).
\newblock \href {http://arxiv.org/abs/1909.12181} {\path{arXiv:1909.12181}}.

\bibitem{Maldacena:1997re}
J.~M. Maldacena, {The Large N limit of superconformal field theories and
  supergravity}, Adv. Theor. Math. Phys. 2 (1998) 231--252.
\newblock \href {http://arxiv.org/abs/hep-th/9711200}
  {\path{arXiv:hep-th/9711200}}, \href
  {https://doi.org/10.4310/ATMP.1998.v2.n2.a1}
  {\path{doi:10.4310/ATMP.1998.v2.n2.a1}}.

\bibitem{Karch:2000gx}
A.~Karch, L.~Randall, {Open and closed string interpretation of SUSY CFT's on
  branes with boundaries}, JHEP 06 (2001) 063.
\newblock \href {http://arxiv.org/abs/hep-th/0105132}
  {\path{arXiv:hep-th/0105132}}, \href
  {https://doi.org/10.1088/1126-6708/2001/06/063}
  {\path{doi:10.1088/1126-6708/2001/06/063}}.

\bibitem{Buhl-Mortensen:2016jqo}
I.~Buhl-Mortensen, M.~de~Leeuw, A.~C. Ipsen, C.~Kristjansen, M.~Wilhelm, {A
  Quantum Check of AdS/dCFT}, JHEP 01 (2017) 098.
\newblock \href {http://arxiv.org/abs/1611.04603} {\path{arXiv:1611.04603}},
  \href {https://doi.org/10.1007/JHEP01(2017)098}
  {\path{doi:10.1007/JHEP01(2017)098}}.

\bibitem{Hoppe:1982phd}
J.~Hoppe, {Quantum theory of a massless relativistic surface and a
  two-dimensional bound state problem}, Ph.D. thesis, {Massachusetts Institute
  of Technology, Dept. of Physics} ({1982}).

\bibitem{Madore:1991bw}
J.~Madore, {The Fuzzy sphere}, Class. Quant. Grav. 9 (1992) 69--88.
\newblock \href {https://doi.org/10.1088/0264-9381/9/1/008}
  {\path{doi:10.1088/0264-9381/9/1/008}}.

\bibitem{Herzog:2015ioa}
C.~P. Herzog, K.-W. Huang, K.~Jensen, {Universal Entanglement and Boundary
  Geometry in Conformal Field Theory}, JHEP 01 (2016) 162.
\newblock \href {http://arxiv.org/abs/1510.00021} {\path{arXiv:1510.00021}},
  \href {https://doi.org/10.1007/JHEP01(2016)162}
  {\path{doi:10.1007/JHEP01(2016)162}}.

\bibitem{Herzog:2017xha}
C.~P. Herzog, K.-W. Huang, {Boundary Conformal Field Theory and a Boundary
  Central Charge}, JHEP 10 (2017) 189.
\newblock \href {http://arxiv.org/abs/1707.06224} {\path{arXiv:1707.06224}},
  \href {https://doi.org/10.1007/JHEP10(2017)189}
  {\path{doi:10.1007/JHEP10(2017)189}}.

\bibitem{Herzog:2017kkj}
C.~Herzog, K.-W. Huang, K.~Jensen, {Displacement Operators and Constraints on
  Boundary Central Charges}, Phys. Rev. Lett. 120~(2) (2018) 021601.
\newblock \href {http://arxiv.org/abs/1709.07431} {\path{arXiv:1709.07431}},
  \href {https://doi.org/10.1103/PhysRevLett.120.021601}
  {\path{doi:10.1103/PhysRevLett.120.021601}}.

\bibitem{Chalabi:2021jud}
A.~Chalabi, C.~P. Herzog, A.~O'Bannon, B.~Robinson, J.~Sisti, {Weyl anomalies
  of four dimensional conformal boundaries and defects}, JHEP 02 (2022) 166.
\newblock \href {http://arxiv.org/abs/2111.14713} {\path{arXiv:2111.14713}},
  \href {https://doi.org/10.1007/JHEP02(2022)166}
  {\path{doi:10.1007/JHEP02(2022)166}}.

\bibitem{Drukker:2022pxk}
N.~Drukker, Z.~Kong, G.~Sakkas, {Broken Global Symmetries and Defect Conformal
  Manifolds}, Phys. Rev. Lett. 129~(20) (2022) 201603.
\newblock \href {http://arxiv.org/abs/2203.17157} {\path{arXiv:2203.17157}},
  \href {https://doi.org/10.1103/PhysRevLett.129.201603}
  {\path{doi:10.1103/PhysRevLett.129.201603}}.

\bibitem{Herzog:2023dop}
C.~P. Herzog, V.~Schaub, {The Tilting Space of Boundary Conformal Field
  Theories} (1 2023).
\newblock \href {http://arxiv.org/abs/2301.10789} {\path{arXiv:2301.10789}}.

\bibitem{deLeeuw:2023wjq}
M.~de~Leeuw, C.~Kristjansen, G.~Linardopoulos, M.~Volk, {B-type anomaly
  coefficients for the D3-D5 domain wall}, Phys. Lett. B 846 (2023) 138235.
\newblock \href {http://arxiv.org/abs/2307.10946} {\path{arXiv:2307.10946}},
  \href {https://doi.org/10.1016/j.physletb.2023.138235}
  {\path{doi:10.1016/j.physletb.2023.138235}}.

\bibitem{Bianchi:2022ppi}
L.~Bianchi, D.~Bonomi, {Conformal dispersion relations for defects and
  boundaries}, SciPost Phys. 15~(2) (2023) 055.
\newblock \href {http://arxiv.org/abs/2205.09775} {\path{arXiv:2205.09775}},
  \href {https://doi.org/10.21468/SciPostPhys.15.2.055}
  {\path{doi:10.21468/SciPostPhys.15.2.055}}.

\bibitem{Bissi:2018mcq}
A.~Bissi, T.~Hansen, A.~S\"oderberg, {Analytic Bootstrap for Boundary CFT},
  JHEP 01 (2019) 010.
\newblock \href {http://arxiv.org/abs/1808.08155} {\path{arXiv:1808.08155}},
  \href {https://doi.org/10.1007/JHEP01(2019)010}
  {\path{doi:10.1007/JHEP01(2019)010}}.

\bibitem{Gimenez-Grau:2020jvf}
A.~Gimenez-Grau, P.~Liendo, P.~van Vliet, {Superconformal boundaries in
  $4-\epsilon$ dimensions}, JHEP 04 (2021) 167.
\newblock \href {http://arxiv.org/abs/2012.00018} {\path{arXiv:2012.00018}},
  \href {https://doi.org/10.1007/JHEP04(2021)167}
  {\path{doi:10.1007/JHEP04(2021)167}}.

\bibitem{Lauria:2020emq}
E.~Lauria, P.~Liendo, B.~C. Van~Rees, X.~Zhao, {Line and surface defects for
  the free scalar field}, JHEP 01 (2021) 060.
\newblock \href {http://arxiv.org/abs/2005.02413} {\path{arXiv:2005.02413}},
  \href {https://doi.org/10.1007/JHEP01(2021)060}
  {\path{doi:10.1007/JHEP01(2021)060}}.

\bibitem{Behan:2020nsf}
C.~Behan, L.~Di~Pietro, E.~Lauria, B.~C. Van~Rees, {Bootstrapping
  boundary-localized interactions}, JHEP 12 (2020) 182.
\newblock \href {http://arxiv.org/abs/2009.03336} {\path{arXiv:2009.03336}},
  \href {https://doi.org/10.1007/JHEP12(2020)182}
  {\path{doi:10.1007/JHEP12(2020)182}}.

\bibitem{Behan:2021tcn}
C.~Behan, L.~Di~Pietro, E.~Lauria, B.~C. van Rees, {Bootstrapping
  boundary-localized interactions II. Minimal models at the boundary}, JHEP 03
  (2022) 146.
\newblock \href {http://arxiv.org/abs/2111.04747} {\path{arXiv:2111.04747}},
  \href {https://doi.org/10.1007/JHEP03(2022)146}
  {\path{doi:10.1007/JHEP03(2022)146}}.

\bibitem{Minahan:2002ve}
J.~A. Minahan, K.~Zarembo, {The Bethe ansatz for N=4 superYang-Mills}, JHEP 03
  (2003) 013.
\newblock \href {http://arxiv.org/abs/hep-th/0212208}
  {\path{arXiv:hep-th/0212208}}, \href
  {https://doi.org/10.1088/1126-6708/2003/03/013}
  {\path{doi:10.1088/1126-6708/2003/03/013}}.

\bibitem{Basso:2015zoa}
B.~Basso, S.~Komatsu, P.~Vieira, {Structure Constants and Integrable Bootstrap
  in Planar N=4 SYM Theory} (5 2015).
\newblock \href {http://arxiv.org/abs/1505.06745} {\path{arXiv:1505.06745}}.

\bibitem{Komatsu:2017buu}
S.~Komatsu, {Three-point functions in $\mathcal N=$ 4 supersymmetric
  Yang\textendash{}Mills theory} (10 2017).
\newblock \href {http://arxiv.org/abs/1710.03853} {\path{arXiv:1710.03853}},
  \href {https://doi.org/10.1093/oso/9780198828150.003.0010}
  {\path{doi:10.1093/oso/9780198828150.003.0010}}.

\bibitem{Georgiou:2023yak}
G.~Georgiou, G.~Linardopoulos, D.~Zoakos, {Holographic correlators of
  semiclassical states in defect CFTs}, Phys. Rev. D 108~(4) (2023) 046016.
\newblock \href {http://arxiv.org/abs/2304.10434} {\path{arXiv:2304.10434}},
  \href {https://doi.org/10.1103/PhysRevD.108.046016}
  {\path{doi:10.1103/PhysRevD.108.046016}}.

\bibitem{Gimenez-Grau:2023fcy}
A.~Gimenez-Grau, {The Witten Diagram Bootstrap for Holographic Defects} (6
  2023).
\newblock \href {http://arxiv.org/abs/2306.11896} {\path{arXiv:2306.11896}}.

\bibitem{Aharony:2008ug}
O.~Aharony, O.~Bergman, D.~L. Jafferis, J.~Maldacena, {N=6 superconformal
  Chern-Simons-matter theories, M2-branes and their gravity duals}, JHEP 10
  (2008) 091.
\newblock \href {http://arxiv.org/abs/0806.1218} {\path{arXiv:0806.1218}},
  \href {https://doi.org/10.1088/1126-6708/2008/10/091}
  {\path{doi:10.1088/1126-6708/2008/10/091}}.

\bibitem{Minahan:2008hf}
J.~A. Minahan, K.~Zarembo, {The Bethe ansatz for superconformal Chern-Simons},
  JHEP 09 (2008) 040.
\newblock \href {http://arxiv.org/abs/0806.3951} {\path{arXiv:0806.3951}},
  \href {https://doi.org/10.1088/1126-6708/2008/09/040}
  {\path{doi:10.1088/1126-6708/2008/09/040}}.

\bibitem{Kristjansen:2021abc}
C.~Kristjansen, D.-L. Vu, K.~Zarembo, {Integrable domain walls in ABJM theory},
  JHEP 02 (2022) 070.
\newblock \href {http://arxiv.org/abs/2112.10438} {\path{arXiv:2112.10438}},
  \href {https://doi.org/10.1007/JHEP02(2022)070}
  {\path{doi:10.1007/JHEP02(2022)070}}.

\bibitem{Gombor:2022aqj}
T.~Gombor, C.~Kristjansen, {Overlaps for matrix product states of arbitrary
  bond dimension in ABJM theory}, Phys. Lett. B 834 (2022) 137428.
\newblock \href {http://arxiv.org/abs/2207.06866} {\path{arXiv:2207.06866}},
  \href {https://doi.org/10.1016/j.physletb.2022.137428}
  {\path{doi:10.1016/j.physletb.2022.137428}}.

\end{thebibliography}

\end{document}